\documentclass[aps,twocolumn,superscriptaddress,unsortedaddress,floatfix,amsmath,amssymb,showpacs]{revtex4}
\usepackage{graphicx}
\usepackage{dcolumn}
\usepackage{bm}
\begin{document}

\title {Optical properties of coupled metal-semiconductor and metal-molecule nanocrystal complexes:
the role of multipole effects}

\author{Jie-Yun Yan}
\affiliation{Institute of Applied Physics and Computational
Mathematics, P. O. Box 8009, Beijing 100088, China}
\affiliation{Department of Physics, Tsinghua University, Beijing
100084, China}
\author{Wei Zhang* } \affiliation{Institute of Applied
Physics and Computational Mathematics, P. O. Box 8009, Beijing
100088, China}
\author{Suqing Duan}
\affiliation{Institute of Applied Physics and Computational
Mathematics, P. O. Box 8009, Beijing 100088, China}
\author{Xian-Geng Zhao}
\affiliation{Institute of Applied Physics and Computational
Mathematics, P. O. Box 8009, Beijing 100088, China}
\author{Alexander O. Govorov **} \affiliation{Department of Physics
and Astronomy, Ohio University, Athens, OH 45701-2979, USA}

\begin{abstract}
We investigate theoretically the effects of interaction between an
optical dipole (semiconductor quantum dot or molecule) and metal
nanoparticles. The calculated absorption spectra of hybrid
structures demonstrate strong effects of interference coming from
the exciton-plasmon coupling. In particular, the absorption
spectra acquire characteristic asymmetric lineshapes and strong
anti-resonances. We present here an exact solution of the problem
beyond the dipole approximation and find that the multipole
treatment of the interaction is crucial for the understanding of
strongly-interacting exciton-plasmon nano-systems. Interestingly,
the visibility of the exciton resonance becomes greatly enhanced
for small inter-particle distances due to the interference
phenomenon, multipole effects, and electromagnetic enhancement. We
find that the destructive interference is particularly strong.
Using our exact theory, we show that the interference effects can
be observed experimentally even in the exciting systems at room
temperature.
\end{abstract}


\maketitle

\section{Introduction}

Nanomaterials incorporating semiconductor quantum dots (SQDs), metal
nanoparticles (MNPs), metal surfaces, and dye molecules have been
studied intensively
\cite{Cui2001,Yin2005,Lakowicz,Shimizu,Anger,Lee,Liu,Ito,Slocik,
Govorov, Persson, Dung,Edwards}.  Such hybrid structures take
advantage of physical properties of different material systems and
demonstrate useful sensor and light-harvesting properties.  In a
complex made of emitter (SQD or dye) and MNPs, excitons and plasmons
interact via the Coulomb forces. Emission of SQD (dye) in the
presence of MNPs can be suppressed or enhanced depending on the size
and organization of nano-assembly \cite{Lakowicz,Anger,Govorov}.
This occurs due to the following physical factors: (1) modified
density of states of photons, (2) amplified absorption in the
presence of plasmon resonance in the MNPs, and (3) shortening of
exciton lifetime due to an increased radiation rate and energy
transfer to the MNPs. Theoretically, an emitting dipole in the
vicinity of metal objects has been studied in many publications
\cite{Lakowicz,Anger,Govorov,Persson,Dung,Klimov}. The
exciton-plasmon interaction between an emitting dipole and metal
objects leads to energy transfer, a shift of energy of quantum
emitter, and vacuum Rabi oscillations
\cite{Govorov,Persson,Dung,Klimov}. The energy transfer mechanism is
similar to F$\mathbf{\ddot{o}}$rster transfer between two dye
molecules \cite{Foster}. In most theoretical publications, the
energy transfer in nanoscale systems was treated as uni-directional
flow of energy from a donor to accepter using the rate equations.
This rate-equation approach describes non-coherent interactions at
elevated temperatures. Very recently, several theoretical studies
were performed for the quantum regime of exciton-plasmon interaction
between a SQD/dye and MNP \cite{Zhang2006,Kelley,Neuhauser,Cheng}.
These studies revealed a novel feature of strongly-interacting
hybrid nanocrystals - interference effects. The absorption spectra
of coupled SQD and MNP acquire characteristic asymmetry due to the
interference of external and induced electric fields. At low
temperatures and small exciton broadening, the absorption line
shapes are described by the Fano formula \cite{Fano}.   However,
previous theoretical results \cite{Zhang2006,Kelley,Neuhauser,Cheng}
were based on the dipole approximation. The dipole approximation is
valid when an exciton-MNP distance is large, compared to the sizes
of components, and the exciton-plasmon interaction is relatively
weak.  But, the most interesting regime of strong exciton-plasmon
interaction appears when a exciton-MNP distance is small and the
dipole approximation is not valid anymore. Therefore, to describe
the regime of strong exciton-plasmon coupling, one should treat the
Coulomb interaction exactly, including electric multipole effects.
In this paper we obtain an exact solution for the problem of
interacting dipole and MNP.

Here we present a theory of strong exciton-plasmon interaction and
show that the multipole effects are of crucial importance for the
understanding of the exciton-plasmon interaction in the most
interesting regime of small exciton-MNP separations. A strong
interaction between excitons and plasmons at small exciton-MNP
distances reveals itself in absorption spectra as asymmetric
lineshapes and anti-resonances (deep minima). Moreover, we consider
two types of nanoscale hybrid complexes (SQD-MNP and dye-MNP) and
show that, for small exciton-MNP distances,  the interference
effects can appear in room-temperature experiments. For the
multipole regime of Coulomb interaction, the visibility of exciton
resonance grows dramatically due to both the interference effect and
the plasmon-induced electromagnetic enhancement. For large
inter-particle distances, the absorptions by SQD and MNP add up
constructively. For the hybrid structures of small dimensions, the
spectra exhibit very strong effects of constructive and destructive
interference. We think that these interference effects can be
observed experimentally using presently available material systems.

In this paper, we discuss three material systems: colloidal
nanocrystals, self-assembled dots, and dye molecules. Most of the
current experiments on metal-semiconductor and metal-dye
assemblies employ colloidal nanoparticles \cite{Anger, Lee, Liu,
Edwards, Ito} and nano-wires \cite{Lee}. To observe the effects
described in this paper, single colloidal nanocrystal complexes
can be deposited on a surface or buried in a polymer film.
Assemble measurements of colloidal complexes in a solution can
also be suitable. Presently, there are attempts to fabricated
metal nanoparticles inside epitaxial structures \cite{Driscoll}.
Potentially, colloidal and epitaxial nanocrystals can be combined
in one structure \cite{Woggon}. For example, an epitaxial dot can
be buried close to the surface \cite{Mazur} and a MNP can be
attached to the surface.

The paper is organized as follows. In section II, we give a
description of the electric field inside the system. The optical
properties of a nanocrystal molecule are presented in section III.
Section IV discusses the absorption line shape. The numerical
results for SQD-MNP molecules and dye molecule-MNP nanocrystals
are given in section V. Finally, a brief conclusion is presented.

\section{Electric fields inside the system}
\begin{figure}[tbp]
\includegraphics[width=7.5cm]{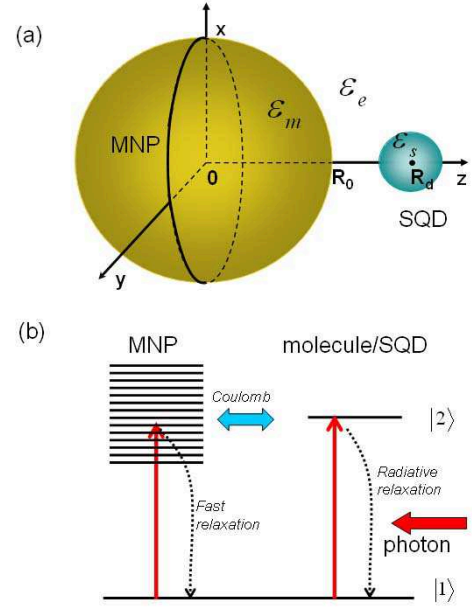}
\caption{a) The model of of semiconductor-netal nanoparticle
molecule. b) The energy structure of the system; the arrows
describe the main transitions.}\label{molecule}
\end{figure}

Now we consider the hybrid nanocrystal molecule composed of a
spherical MNP of radius $R_0$ and a spherical SQD of radius $R_s$ in
the environment with dielectric constant $\varepsilon_e$ (see
Fig.\ref{molecule}(a)). The center-to-center distance for the
nanocrystals is denoted as $R_d$. In the following, we will assume
that a SQD has small dimensions $R_s\ll R_d$, whereas the size of
MNP can be arbitrary, i.e. $R_0\sim R_d$. In the case of a dye
molecule-MNP complex, the size $R_d$ becomes irrelevant and we have
only the obvious condition: the molecule-MNP distance $R_d$ should
be larger than the MNP radius, i.e. $R_d>R_0$.

When the hybrid molecule (SQD-MNP or dye-MNP) is radiated with a
laser, the oscillating electric field excites both the interband
transition in a SQD and the plasmon in a MNP. The interband
transition (exciton) in a SQD (dye) has an oscillating dipole moment
and, therefore, creates an oscillating electric field that acts on a
MNP and excites the plasmons. The plasmons excited in a MNP in turn
influence the exciton. In this way, the exciton and plasmon form a
new collective excitation - a hybrid exciton. The origin of this
hybrid exciton is in the Coulomb interaction between a SQD and MNP
Fig.\ref{molecule}(b).

Since a nanoscale structure has small dimensions, we assume that
the electric field of the laser is spacial uniform and use the
quasistatic approximation. Also, we describe the MNP with the
local dynamic dielectric function $\varepsilon_m(\omega)$.
By carefully solving the Laplace's equations $\nabla^2\varphi=0$,
we can get the electric potential $\varphi$ and so the electric
field $\boldsymbol{E}=-\nabla\varphi$. As for the dipole in SQD,
we take it as a perfect dipole, labeled as $\boldsymbol{p}$,
because the radius of SQD $R_s$ is much smaller in contrast with
$R_0$ usually.

We now consider the laser electric field in the form
$\boldsymbol{E}_0(t)=\tilde{\boldsymbol{E}}_0\cos\omega t$.
We first discuss the part with positive frequency,
$\tilde{\boldsymbol{E}}_0e^{-i\omega t}/2$.
The electric field inside SQD could be split in three parts:
$\boldsymbol{E}_{\texttt{SQD}}=(\tilde{\boldsymbol{E}}_1+\tilde{\boldsymbol{E}}_2
+\tilde{\boldsymbol{E}}_3)e^{-i\omega t}$.
$\tilde{\boldsymbol{E}_1}e^{-i\omega t}$ is the electric field
induced by $\boldsymbol{E}_0$ inside a SQD in the absence of a MNP:
\begin{eqnarray}
\tilde{\boldsymbol{E}}_1=\dfrac{\varepsilon_e}{\varepsilon_{\texttt{eff1}}}
\dfrac{\tilde{\boldsymbol{E}}_0}{2},\label{elec1}
\end{eqnarray}
where $\varepsilon_{\texttt{eff1}}=(\varepsilon_s+2\varepsilon_e)/3$
is the effective dielectric constant of the SQD and $\varepsilon_s$
is the background dielectric constant of semiconductor.

The second part $\tilde{\boldsymbol{E}}_2e^{-i\omega t}$ comes from
surface charges of MNP induced only by the external optical electric
field $\tilde{\boldsymbol{E}}_0 e^{-i\omega t}/2$:
\begin{eqnarray}
\tilde{\boldsymbol{E}}_2
=s_{\alpha}\dfrac{\varepsilon_e\gamma_1R_0^3}{\varepsilon_{\texttt{eff1}}R_d^3}
\dfrac{\tilde{\boldsymbol{E}}_0}{2},\label{elec2}
\end{eqnarray}
where
$\gamma_1=[\varepsilon_m(\omega)-\varepsilon_e]/[\varepsilon_m(\omega)+2\varepsilon_e]$
and $s_{\alpha}=2(-1)$ for electric field polarization
$\boldsymbol{E}_0$ is parallel to the $z(x)$ axis. The choice of a
sign of imaginary part of dialectic constant should be the
following: $\mathrm{Im}[\varepsilon_m(\omega)]>0$ if $\omega>0$.

The last part $\tilde{\boldsymbol{E}}_3e^{-i\omega t}$ is the
SQD-felt effective electric field produced by the multipole
polarization in MNP induced by the effective dipole of SQD
$\boldsymbol{p}=(\varepsilon_e/\varepsilon_{\texttt{eff1}})\tilde{\boldsymbol{p}}e^{-i\omega
t}$. This part is the most important one, underlying the interaction
of the MNP and SQD. We finally find this part of electric field is
\begin{eqnarray}
\tilde{\boldsymbol{E}}_3&=&\sum_{n=1}^{\infty}
\dfrac{s_n\varepsilon_e\gamma_n R_0^{2n+1}}
{\varepsilon_{\texttt{eff1}}^2
R_d^{2n+4}}\tilde{\boldsymbol{p}},\label{elec3}
\end{eqnarray}
where
\begin{eqnarray}
\gamma_n=\dfrac{\varepsilon_m(\omega)-\varepsilon_e}
{\varepsilon_m(\omega)+\frac{n+1}{n}\varepsilon_e},\label{multipole}
\end{eqnarray}
and the coefficients $s_n=(n+1)^2$ or $P'_n(1)$ for the polarization
$\boldsymbol{p}$ is parallel to the $z$ or $x$ axis respectively.
Here $P_n$ is the Legendre function and $P'_n(1)$ is the
differential of Legendre function at the argument of 1.

When $n=1$, the electric field $\tilde{\boldsymbol{E}}_3$ becomes
\begin{eqnarray}
\tilde{\boldsymbol{E}}_3&=& s_1 \dfrac{\gamma_1 \varepsilon_e
R_0^3}{\varepsilon_{\texttt{eff1}}^2 R_d^6}\tilde{\boldsymbol{p}},
\end{eqnarray}
which is the very result under the dipole approximation: the SQD
dipole-induced electric field everywhere inside the MNP is uniform
and equals to the value at the center of MNP. This dipole
approximation is used in Ref.\cite{Zhang2006}, which is reasonable
when the distance $R_d$ is relatively large because the factor
$1/R_d^{2n+4}$ makes other terms negligible. However, when the
distance is comparable to the radius of MNP, i.e. $R_0\sim R_d$,
which is always the case in such molecules, the items $n>1$ become
important and may even have a bigger contribution than the leading
terms. Actually the items with different $n$ are related to
different order multipole polarization in MNP: $n=1$ is dipole,
$n=2$ quadrupole, $n=3$ octopole and so on. The function of
$\gamma_n$, Eqn.(\ref{multipole}), also reflects the
characteristics of the multipole polarization as tells by the Mie
theory \cite{Mie1908,Kelly2003}. But it should notice that the
multipole effect in Mie theory is caused by the comparable radius
of MNP with wavelength of the light. Here the radius is small
enough to neglect the effect and hence the multipole effect
totally comes from the induced polarization by the SQD when the
the interparticle distance is relatively small.

The electric field inside MNP $\boldsymbol{E}_{\texttt{MNP,tot}}$
can also be solved, and the corresponding positive frequency part is
$\tilde{\boldsymbol{E}}_{\texttt{MNP,tot}}e^{-i\omega t}=
(\frac{\varepsilon_e}{\varepsilon_{\texttt{eff2}}}\frac{\tilde{\boldsymbol{E}}_0}{2}+
\tilde{\boldsymbol{E}}_{\texttt{MNP}})e^{-i\omega t}$, where
$\varepsilon_{\texttt{eff2}}=[\varepsilon_m(\omega)+2\varepsilon_e]/3$
and $\tilde{\boldsymbol{E}}_{\texttt{MNP}}$ is
\begin{eqnarray}
&&\tilde{\boldsymbol{E}}_{\texttt{MNP}}|_{\boldsymbol{E}\parallel\hat{\boldsymbol{z}}}
=\sum_{n=1}^{\infty}\dfrac{(n+1)\beta_nR^{n-1}|\tilde{\boldsymbol{p}}|}{\varepsilon_{\texttt{eff1}}R_d^{n+2}}\nonumber\\
&&\qquad\times\left[-\frac{x\hat{\boldsymbol{x}}+y\hat{\boldsymbol{y}}}{R}P'_{n-1}(z/R)
+nP_{n-1}(z/R)\hat{\boldsymbol{z}}\right]\label{MNP1}\\
&&\tilde{\boldsymbol{E}}_{\texttt{MNP}}|_{\boldsymbol{E}\parallel\hat{\boldsymbol{x}}}
=\sum_{n=1}^{\infty}\dfrac{\beta_nR^{n-1}|\tilde{\boldsymbol{p}}|}
{\varepsilon_{\texttt{eff1}}R_d^{n+2}}\bigg\{\Big[-nP_{n-1}(z/R)\nonumber\\
&&\qquad-\frac{z}{R}P'_{n-1}(z/R)+\frac{x^2}{R^2}
P''_{n-1}(z/R)\Big]\hat{\boldsymbol{x}}\nonumber\\
&&\qquad+\dfrac{xy}{R^2}P''_{n-1}(z/R)\hat{\boldsymbol{y}}-(n+1)\dfrac{x}{R}P'_{n-1}(z/R)
\hat{\boldsymbol{z}}\bigg\}\label{MNP2}
\end{eqnarray}
where
\begin{eqnarray}
\beta_n=\dfrac{\frac{2n+1}{n}\varepsilon_e}{\epsilon_m(\omega)+\frac{n+1}{n}\varepsilon_e}.\label{MNP3}
\end{eqnarray}
When $n=1$, the electric field inside the MNP,
\begin{eqnarray}
\tilde{\boldsymbol{E}}_{\texttt{MNP}}=s_\alpha\dfrac{1}{\varepsilon_\texttt{eff2}R_d^3}
\left(\dfrac{\varepsilon_e}{\varepsilon_{\texttt{eff1}}}\tilde{\boldsymbol{p}}\right),\label{MNP4}
\end{eqnarray}
is just the electric field produced by the dipole inside the SQD at
the point of MNP's spherical center, the dipole approximation.

\section{Density matrix of the system}
In the SQD, we use a two-level exciton model to describe the optical
process. The exciting laser field is now taken in the form:
$\boldsymbol{E}_0(t)=\tilde{\boldsymbol{E}}_0\cos(\omega t)$. Under
these conditions, the Hamiltonian is
\begin{eqnarray}
H=\hbar\omega_0 a_2^{\dag}a_2-\chi(t)(a_2^{\dag}a_1+a_1^{\dag}a_2),
\end{eqnarray}
where $a_i$ and $a_i^{\dag}$ is the annihilation and creation
operator of level $i$ ($i=1,2$), corresponding to the vacuum ground
state and the exciton state respectively (see
Fig.\ref{molecule}(b)). Note that the energy of level $1$ is taken
as zero. $\chi(t)=\tilde{\chi}e^{-i\omega t}+\tilde{\chi}^*e^{
i\omega
t}=\boldsymbol{\mu}\cdot({\boldsymbol{E}}_{\texttt{SQD}}+{\boldsymbol{E}}^*_{\texttt{SQD}})$,
and $\boldsymbol{\mu}$ is the interband optical transition matrix
element. The excitation $\tilde{\chi}$ can be written as:
$\tilde{\chi}=\boldsymbol{\mu}\cdot(\tilde{\boldsymbol{E}}_1+\tilde{\boldsymbol{E}}_2
+\tilde{\boldsymbol{E}}_3)$; the terms in this equation are defined
by Eqns.(\ref{elec1}, \ref{elec2} and \ref{elec3}).

These matrix elements satisfy the well-known optical Bloch
equations:
\begin{eqnarray}
&\partial_t{\rho}_{22}=\dfrac{i}{\hbar}\chi(t)\left[\rho_{21}^*-\rho_{21}\right]-\dfrac{1}{T_1}\rho_{22},\\
&\partial_t \rho_{21}=-i\omega_0
\rho_{21}+\dfrac{i\chi(t)}{\hbar}(1-2\rho_{22})-\dfrac{1}{T_{20}}\rho_{21}.
\end{eqnarray}
Also, $\rho_{11}+\rho_{22}=1$. $T_1$ and $T_{20}$ are the
longitudinal and transverse dephasing times. The dipole of the SQD
relates to the interband polarization by
$\boldsymbol{p}=\boldsymbol{\mu}(\rho_{21}+\rho_{21}^*)$,
$\rho_{21}=\langle a_1^{\dag}a_2\rangle$. In the rotating wave and
steady state approximation,
$\boldsymbol{p}=\tilde{\boldsymbol{p}}e^{-i\omega
t}+\tilde{\boldsymbol{p}}^*e^{i\omega t}$, where
$\tilde{\boldsymbol{p}}$ is time-independent. Then the equations
give:
\begin{eqnarray}
\rho_{22}=\frac{2T_1}{\hbar}\mathrm{Im}[\tilde{\chi}^*\tilde{\rho_{21}}],\quad
\tilde\rho_{21}=\dfrac{-\Omega\Delta}{(\omega-\omega_0+G\Delta+i/T_{20})},\label{solution}
\end{eqnarray}
where
\begin{eqnarray}
&&\Omega=\dfrac{\boldsymbol{\mu}\cdot(\tilde{\boldsymbol{E}}_1+\tilde{\boldsymbol{E}}_2)}{\hbar},\quad \\
&&G=\sum_{n=1}^{\infty}\dfrac{s_{n}\varepsilon_e\gamma_n
R_0^{2n+1}\mu^2} {\varepsilon^2_{\texttt{eff1}}
R_d^{2n+4}\hbar}\equiv G_R+iG_I,\label{solution2}
\end{eqnarray}
and $G_R$ and $G_I$ are the real and imaginary part of $G$,
$\Delta=\rho_{11}-\rho_{22}$. While $\Omega$ is determined by the
electric field of $\tilde{\boldsymbol{E}}_1$ and
$\tilde{\boldsymbol{E}}_2$, $G$ comes totally from the contribution
of the electric field $\tilde{\boldsymbol{E}}_3$, responsible for
the interaction between MNP and SQD. Therefore the plasmon-exciton
interaction leads to the formation of a hybrid exciton with shifted
exciton frequency and decreased lifetime determined by $G_R$ and
$G_I$ respectively.

\begin{figure}[tbp]
\includegraphics[width=8.5cm]{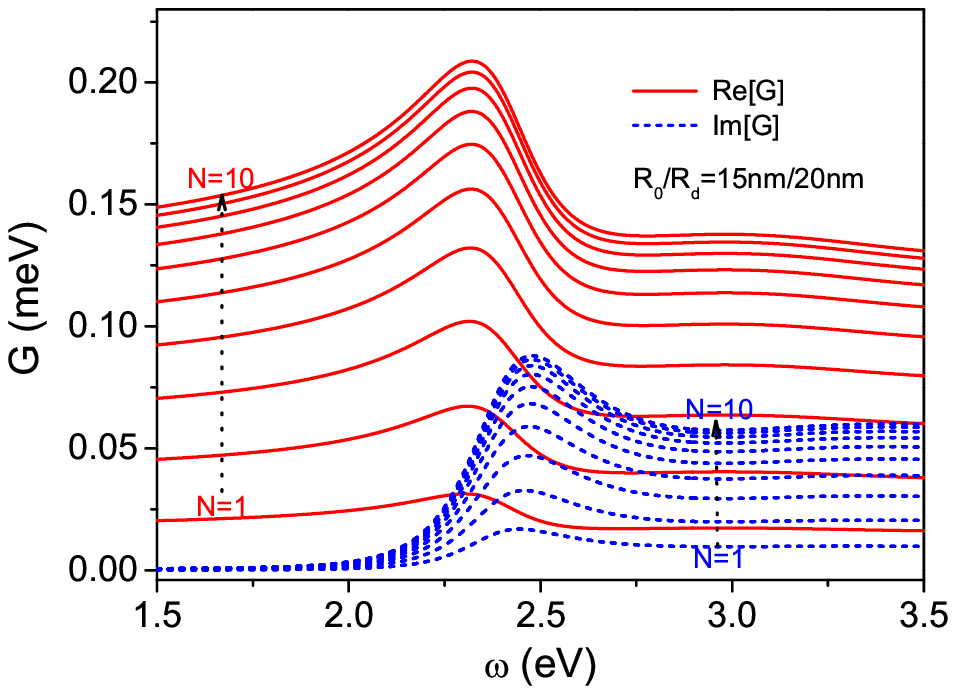}
\includegraphics[width=8.5cm]{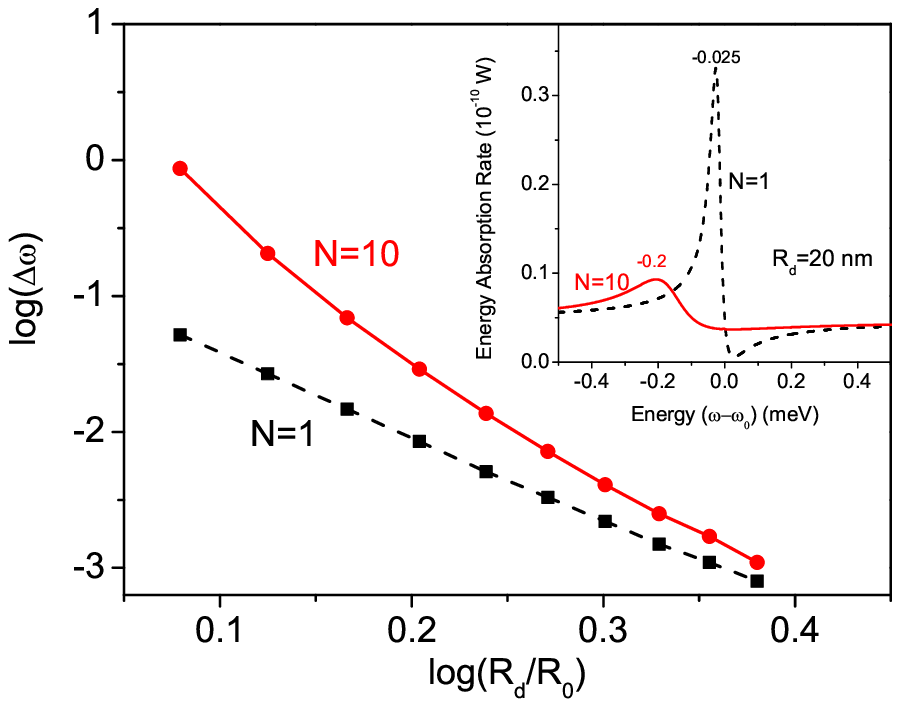}
\caption{Top: The demonstration of multipole effect in system with
MNP of Au. Solid (Red) lines and dash (Blue) lines represents the
real and imaginary part of $G$ respectively. From bottom to top, $N$
changes from 1 to 10. Radius of MNP and the distance is 15 nm and 20
nm respectively. Bottom: Peak shift of energy absorption rate with
different interparticle distances in logarithmic coordinate. The
solid (Red) line and circle is the one including multipole effect
($N=10$), while the dash (Black) line and square is only in dipole
approximation ($N=1$). Inset is contrast of energy absorption rate
for a typical interparticle distances $R_d$=20 nm, with peak value
emphasized. }\label{AuG}
\end{figure}

For a weak external field,
$\Delta=\rho_{11}-\rho_{22}=1-2\rho_{22}\approx 1$ (since
$\rho_{22}\ll1$) and Eqn.(\ref{solution}) and Eqn.(\ref{solution2})
yield the solution of the problem. For a strong driving field, the
problem is reduced to the nonlinear equation (Eqn.(\ref{solution}))
investigated in ~\cite{Zhang2006}. Since the quantity $G_R$ includes
multipole polarization in MNP, the energy shift is determined by the
multipole effect rather than a simple dipole effect. In
Fig.\ref{AuG}(top), we plot the $G=\sum_{n=1}^N(\cdots)$ with the
change of frequency $\omega$. The MNP is made of gold and dielectric
constant is taken from Ref.\cite{Johnson1972}. Other parameters are
chosen as: $\varepsilon_s=6$,
$R_0$=15 nm, $R_d$=20 nm. The chosen SQD parameters are typical
for the CdSe and CdTe colloidal quantum-dot systems.
It shows that the excitonic frequency shift $G_R$ enhances with $N$
increasing and finally converges. Due to the factor $1/R_d^{2n+4}$,
the terms of higher order have smaller contribution. However, in
finite range, the contribution from terms of $n>1$ is more than that
of $n=1$. The shift when $N=10$ is almost seven times than that of
$N=1$. As for the lifetime of the exciton, it decreases with $N$
increasing due to $G_I$. More close the distance of the molecule,
more multipole contribution should be considered. In some case such
as the frequency $\omega$ locates at the value making
$Re[\varepsilon_m(\omega)+3/2\varepsilon_e]=0$, the quadrupole
effect becomes the leading one. As for the lifetime of exciton, the
interaction makes it shorter and multipole interaction boosts up
this effect. For Au MNP, the lifetime is almost unchanged when
$\omega$ below 2 eV while the exciton frequency shift is still
outstanding, which may has important applications.

The multipole effects can also be seen in the energy absorption
spectrum. Under the radiation of laser, both SQD and MNP absorb
energy, which will eventually transformed into heat by all kinds of
scattering mechanics. The energy absorption rate is actually the
time-average of power dissipation density. In SQD, the energy
absorption rate $Q_{\texttt{SQD}}$ is
$Q_{\texttt{SQD}}=\hbar\omega_0\rho_{22}/T_1$.

As for the MNP, the optical transmission property is more measured
by the quantity of extinction, which has two main parts: absorption
and scattering. The relative contribution of scattering to the
extinction for metal particles of radius less than about 30 nm can
be negligible. In the MNP-SQD molecules fabricated in recent
experiments, the size of MNP is within the range to discount the
scattering. For example, the Rayleigh scattering has been calculated
and found about three order of magnitude smaller than the energy
absorption rate\cite{Zhang2006}. The energy absorption rate in MNP
$Q_{\texttt{MNP}}$ is expressed as
$Q_{\texttt{MNP}}=\frac{\omega}{2\pi}\int
\mathrm{Im}[\varepsilon_m(\omega)]|\tilde{\boldsymbol{E}}_{\texttt{MNP,tot}}|^2
dV$, where $\tilde{\boldsymbol{E}}_{\texttt{MNP,tot}}=
\frac{\varepsilon_e}{\varepsilon_{\texttt{eff2}}}\frac{\tilde{\boldsymbol{E}}_0}{2}+
\tilde{\boldsymbol{E}}_{\texttt{MNP}}$ and
$\tilde{\boldsymbol{E}}_{\texttt{MNP}}$ is given by
Eqns.(\ref{MNP1}-\ref{MNP4}) The total rate of absorption is
therefore:
\begin{eqnarray}
Q_{\texttt{tot}}=Q_{\texttt{MNP}}+Q_{\texttt{SQD}}.\label{absorb}
\end{eqnarray}

A typical energy absorption spectrum is shown in the insert of
Fig.\ref{AuG}(Bottom).  Here the energy absorption rate including
multipole effect for a typical value $R_d$=20 nm is plotted in
contrast with the result in dipole approximation in the inset. The
interaction of the discrete state in SQD and the continuous states
in MNP gives birth to the Fano lineshape. The laser intensity is
$I_0=1$ W/cm$^2$. The original excitonic energy is
$\hbar\omega_0$=2.5 eV, the plasmon peak of Au. Fig.\ref{AuG} also
plots peak shift of energy absorption rate with different
interparticle distances in weak field regime in logarithmic
coordinate. With the distance decreasing, the spectra has peak
shifts and broadening. In dipole approximation, the slope of peak
shift is -6 for $G\sim 1/R_d^6$; The multipole effect accounts for
significant change in the relation of $G$ and $R_d$ and leads to
larger shift of peak. Especially when the value of $R_d/R_0$ is
small, the shift increases almost by one order of magnitude, which
makes it more possible to be observed experimentally. As the
interdistance of the molecule can be adjusted through biolinker by
temperature, the shift is temperature-controlled, which may have
important application in sensor and detector. The broadening of
spectra is related to the incoherent energy transfer rate. The
incoherent energy transfer in the hybrid molecule is via the Forster
mechanism with energy transfer rate $G_I$. As the multipole effect
increase the value of $G_I$ in the vicinity of the plasmon peak, see
Fig.\ref{AuG}, the energy transfer time changes remarkably.

\begin{figure}[tbp]
\includegraphics[width=8.5cm]{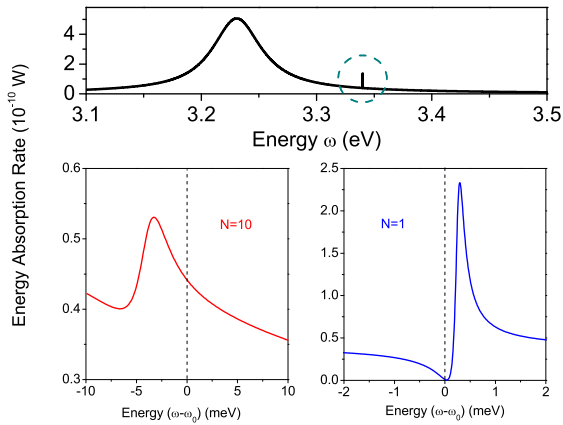}
\caption{The different shift caused by the multipole effect in
contrast with dipole approximation. The Ag MNP and SQD is emerged in
water ($\varepsilon_0=1.8$). Other parameters are
$R_d$=22 nm, $R_0$=15 nm and $\hbar\omega_0$=3.34 eV.}\label{shift}
\end{figure}

Multipole effects not only lead to  quantitative changes (which is
important for experiments),
 but also  qualitative difference. For example, we
consider the Ag MNP and SQD composed molecule emerged in water
solution. As shown in Fig.\ref{shift}, in dipole approximation, the
peak of absorption rate has blue shift. With multipole effect, the
peak's shift of energy absorption rate changes to the red direction
and the broadening increases a lot.

\begin{figure}[tbp]
\includegraphics[width=8.5cm]{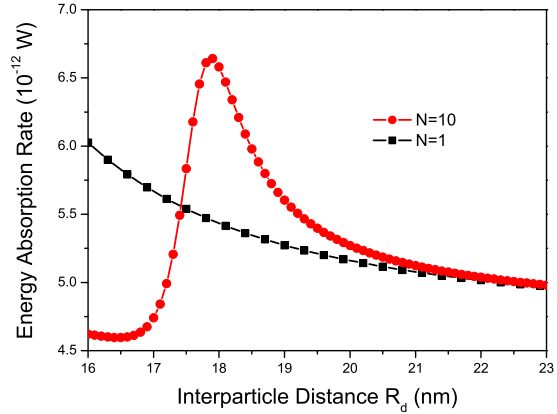}
\caption{Total energy absorption versus interparticle distances for
a fixed frequency $\hbar\omega=2.499$ eV. $R_{\texttt{MNP}}=15$ nm.
Other parameters are the same as those for figure
\ref{AuG}.}\label{changRD}
\end{figure}

The total energy absorption versus interparticle distances for a
fixed frequency is presented in Fig.\ref{changRD}, which shows that
with decreasing the interparticle distance, the energy absorption
first increases, then decreases. That means there exists an optimal
distance for energy absorption. Here the dipole approximation fails
to provide the correct picture. The energy absorption of MNP and SQD
versus interparticle distance show the similar results (not shown
here). The existence of optimal distance for SQD energy absorption
can be understood in the following way. The energy absorption of SQD
depends on the competition of two factors: the local field
enhancement and transfer the energy to MNP. With decreasing
interparticle distance, local field becomes larger and energy
transfer rate increases also.
 At short distance, the factor from fast energy transfer wins over the factor
from local field enhancement. In fact Forster transfer (FT) rate is
$1/\delta^3$, here $\delta$ is the  distance between the exciton and
MNP surface ($\delta=R_d-R_0$)\cite{Persson}. When
$\delta\rightarrow0$, the FT rate $\rightarrow \infty$.

\section{Absorption line shapes}
Using Eqn.(\ref{solution}), Eqn.(\ref{absorb}), we can write the
absorption rate as
\begin{eqnarray}
Q_{tot}=Q_{\texttt{MNP}}^0+\dfrac{A\bar{\Gamma}_{12}}{(\omega-\bar{\omega}_0)^2+\bar{\Gamma}_{12}^2}
+\dfrac{B(\omega-\bar{\omega}_0)}{(\omega-\bar{\omega}_0)^2+\bar{\Gamma}_{12}^2},
\end{eqnarray}
where $\bar{\Gamma}_{12}=1/T_{20}+G_I$ and
$\bar{\omega}_0=\omega_0+G_R$ are the renormalized off-diagonal
broadening and the modified absorption frequency, respectively.
The absorption by an isolated MNP is given by
$Q_{\texttt{MNP}}^0=\frac{\omega}{8\pi}\mathrm{Im}[\varepsilon_m(\omega)]
|\frac{\varepsilon_eE_0}{\varepsilon_{\texttt{eff2}}}|^2V_{\texttt{MNP}}$,
where $V_{\texttt{MNP}}$ is the MNP volume. In the general case,
the coefficients $A$ and $B$ are given by rather complex
expressions. Here it is instructive for us to give these
coefficients in the dipole limit. Then the key function $G$ takes
the form:
$G=\frac{s_1\gamma_1\varepsilon_eR_0^3\mu^2}{\varepsilon_{\texttt{eff1}}^2R_d^6\hbar}$.
Note that $G_I>0$. For the limit $\Gamma_{12}\rightarrow 0$, we
can obtain the Fano resonance \cite{Fano},
\begin{eqnarray}
Q_{\texttt{tot}}=Q_{\texttt{MNP}}^0\dfrac{(\omega-\bar{\omega}_0-q_{\texttt{Fano}}\Delta_{\texttt{int}})^2}
{(\omega-\bar{\omega}_0)^2+\Delta_{\texttt{int}}^2},
\end{eqnarray}
where $\Delta_{\texttt{int}}=G_I$ is the interaction-induced
broadening and
$q_{\texttt{Fano}}(\omega_0,d)=\frac{R_d^3}{s_{\alpha}R_0^3\mathrm{Im}[\gamma_1]}$
is the Fano factor. The line is strongly asymmetric if
$q_{\texttt{Fano}}(\omega_0,R_d)\sim 1$. In the limit
$q_{\texttt{Fano}}(\omega_0,R_d)\gg 1$, the absorption line becomes
close to a Lorentzian. For molecular systems and systems at room
temperature, the off-diagonal decoherence is typically strong:
$\Gamma_{12i}\gg |G|$. In this limit, the coefficients in the
lineshape become:
\begin{eqnarray}
A&=&\dfrac{\omega}{2\hbar}\left(\dfrac{\varepsilon_e\tilde{E}_0\mu}{\varepsilon_{\texttt{eff1}}}\right)^2
\left|1+\dfrac{s_{\alpha}\gamma_1 R_0^3}{R_d^3}\right|\nonumber\\
&&-\tilde{E}_0^2\dfrac{\omega\mu^2 s_1 \varepsilon_e R_0^6 \mathrm{Im}[\gamma_1]} {3\hbar
\varepsilon_{\texttt{eff1}}^2R_d^6}\left|\dfrac{\varepsilon_e}
{\varepsilon_{\texttt{eff2}}}\right|^2\mathrm{Im}[\varepsilon_m(\omega)]\\
B&=&\dfrac{s_{\alpha}\mu^2\varepsilon_e \tilde{E}_0^2\omega
R_0^3}{3\hbar
\varepsilon_{\texttt{eff1}}^2R_d^3}\mathrm{Im}[\varepsilon_m(\omega)]
\left|\dfrac{\varepsilon_e}{\varepsilon_{\texttt{eff2}}}\right|^2\nonumber\\
&&\times\left(1+\dfrac{s_{\alpha}R_0^3\mathrm{Re}[\gamma_1]}{R_d^3}\right)
\end{eqnarray}
In the strong decoherence regime, it is convenient to introduce an
asymmetry factor:
\begin{eqnarray}
&&q_{\texttt{asym}}=\dfrac{A}{B}\nonumber\\
&&=\dfrac{\dfrac{\varepsilon_e}{2}\left|1+\dfrac{s_{\alpha}\gamma_1 R_0^3}{R_d^3}\right|^2
-\dfrac{s_1R_0^6\mathrm{Im}[\gamma_1]}{3R_d^6}\left|\dfrac{\varepsilon_e}{\varepsilon_{\texttt{eff2}}}\right|^2\mathrm{Im}[\varepsilon_m(\omega)]}
{\dfrac{s_{\alpha}R_0^3}{3R_d^3}\left|\dfrac{\varepsilon_e}{\varepsilon_{\texttt{eff2}}}
\right|^2\mathrm{Im}[\varepsilon_m(\omega)]\left(1+\dfrac{s_{\alpha}\mathrm{Re}[\gamma_1]
R_0^3}{R_d^3}\right)}\nonumber
\end{eqnarray}
If $|Q_{\texttt{asym}}|\sim 1$, the lineshape is asymmetric and the
interference effect is strong. We also should note that the
interference effect may appear as a symmetric line with a deep. Such
"antiresonance" may appear if $A<0$ and $|q_{\texttt{asym}}|\gg 1$.
The visibility of the antiresonance can be described by the ratio
$|A/(\bar{\Gamma}_{12} Q^0_{\texttt{MNP}}(\omega_0)|$. Our numerical
simulations show that anti-resonances appear for small
inter-particle distances (see the graphs in the next section).

\section{Numerical results}

In this section, we show the results of numerical calculations of
energy absorptions for various systems. We pay special attention to
the short interparticle distance regime, where the multipole effects
are crucial. We emphasize the important role of the environment,
that is essential to describe and understand experiments. One of the
effects of environment is the dephasing time. Another is the
dielectric constant of a matrix. These parameters has a significant
impact on the asymmetric lineshapes of an absorption spectrum.

\begin{figure}[tbp]
\includegraphics[width=8.5cm]{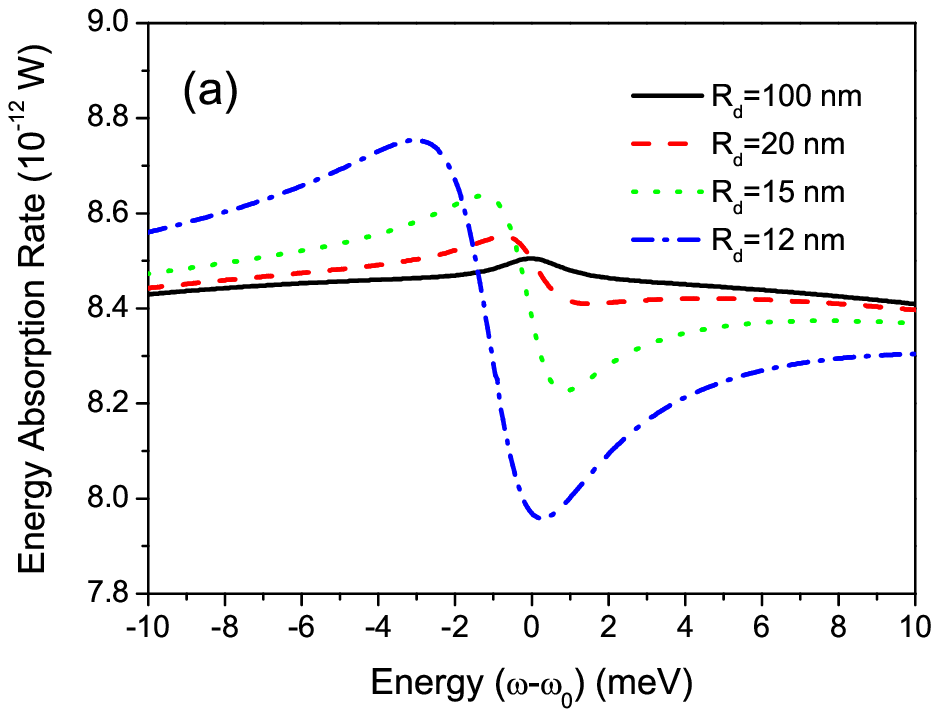}
\includegraphics[width=8.5cm]{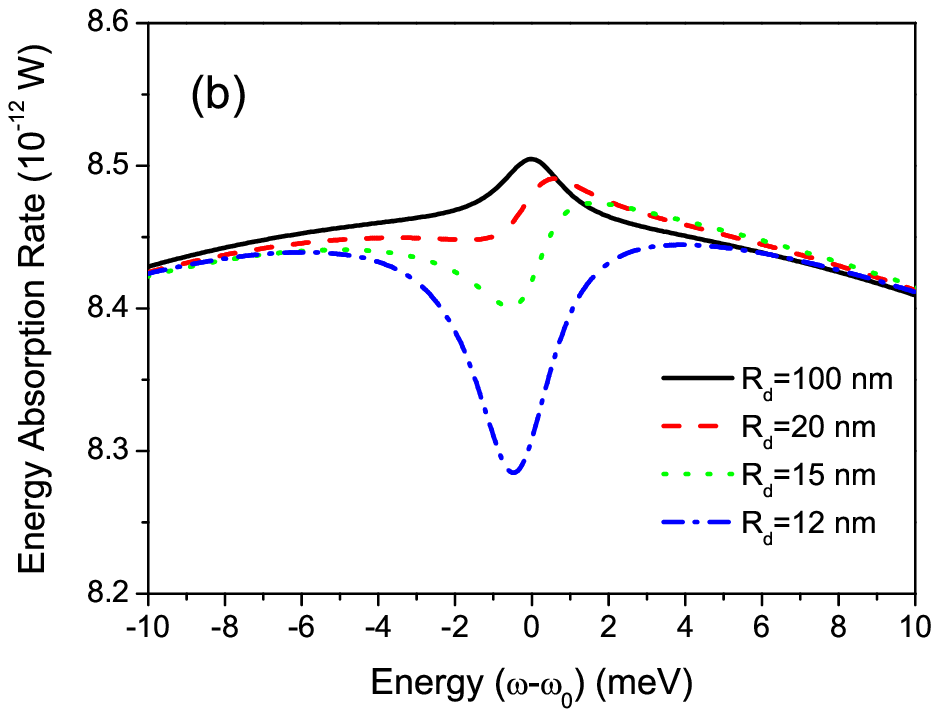}
\includegraphics[width=8.5cm]{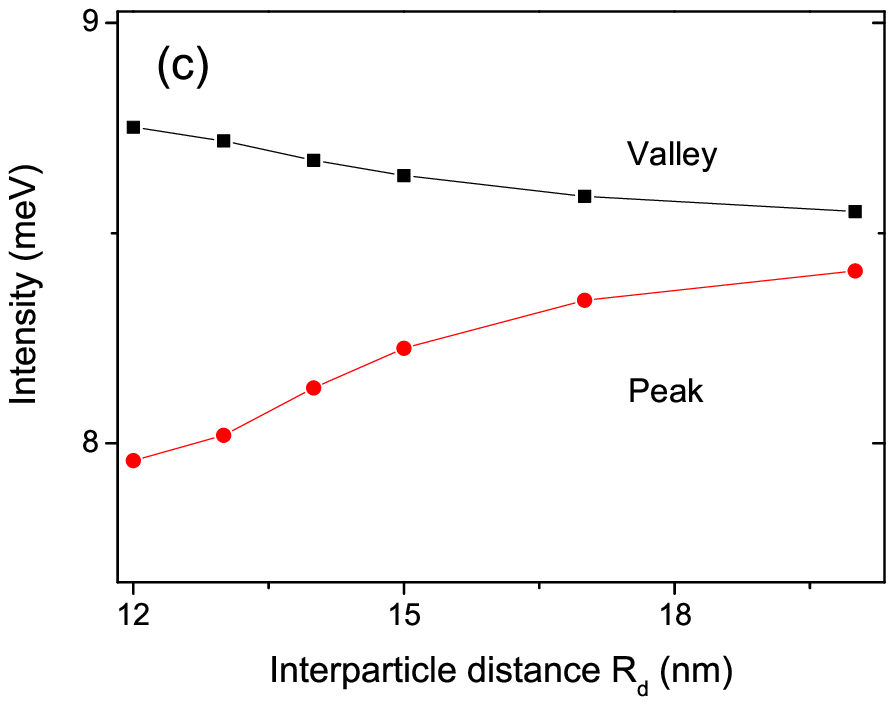}
\caption{The absorption spectrum of SQD-MNP(Au). (a) Polarization is
along z direction. (b) Polarization is along x direction. (c) The
peak and valley intensity versus distance corresponding to (a).
}\label{colloidal}
\end{figure}
\begin{figure}[tbp]
\includegraphics[width=8.5cm]{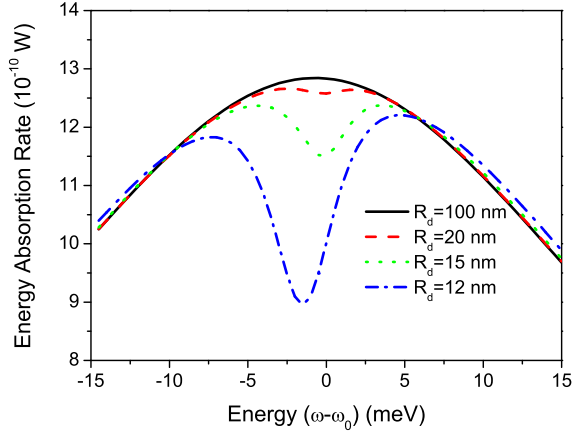}
\caption{The absorption spectrum on epitaxial SQD-MNP(Au).
}\label{epitaxial}
\end{figure}

In Fig.\ref{colloidal}, we show the absorption spectrum of colloidal
SQD-MNP(Au) for different interparticle distances. The parameters
are $1/T_{20}=1$ meV, $T_1=10$ ns, $R_0=10$ nm, $\varepsilon_e=1.8$
(water), $\hbar \omega_0$=2.356 eV. The chosen de-phasing times
correspond to the low-temperature regime when the exciton broadening
$1/T_{20}=1$ meV is relatively small. One can see that the energy
absorption spectrum has a symmetric peak for very large
interparticle distance (100 nm), when the interparticle interaction
is almost absent. With decreasing interparticle distance, an
asymmetric Fano lineshape appears due to the interaction between an
exciton and plasmon \cite{Zhang2006}. From Fig. \ref{colloidal}b, we
can see that with decreasing inter-particle distance, the absorption
spectrum shows quite different behavior for small distance (weak
interaction) and large distance (strong interaction) regime. For
large distance, we observe a blue shifted resonance. While for small
distance, we see a red shifted anti-resonance. The peak/valley
intensity versus distance is show in Fig. \ref{colloidal}c. Overall,
the interference features (asymmetry and anti-resonance) become very
strong at small distances where the multipole effect govern the
inter-particle Coulomb interaction.

It is interesting to note that for large $R_d$ (In
Fig.\ref{colloidal}) the exciton peak and MNP plasmon spectrum are
added constructively whereas for small $R_d$ we obtain destructive
interference or anti-resonance for the laser light along
$\boldsymbol{x}$ or $\boldsymbol{y}$. For the laser light along
$\boldsymbol{z}$, we see both strong constructive and destructive
effects; however, the destructive deep is stronger. The physical
origin of the destructive interference is in a reduction of
absorption by the MNP (i.e. the term $Q_{\texttt{MNP}}$ in the total
absorption) due to the destructive interplay of external and induced
electric fields. Regarding the absorption by the SQD (i.e. the term
$Q_{\texttt{SQD}}\propto\rho_{22}$), the integrated absorption by
the SQD increases with decreasing $R_d$ for
$\tilde{\boldsymbol{E}}_0 || \boldsymbol{z}$ and decreases with
shortening $R_d$ for $\tilde{\boldsymbol{E}}_0 ||
\boldsymbol{x}(\boldsymbol{y})$. The reason is the dynamic screening
of the external field inside the SQD by the MNP. Simultaneously, for
short $R_d$, the exciton peak becomes shifted and strongly
broadened.

In Fig.~\ref{epitaxial}, we show the absorption spectrum of
epitaxial SQD-MNP(Au) for different interparticle distances. The
parameters are $1/T_{20}$=2 meV, $T_1=T_{20}/2$, $R_0=10$ nm,
$\varepsilon_e=\varepsilon_s=12$, the interband dipole matrix
element $\mu=er_0$, $r_0=0.6$ nm, $\hbar \omega_0$=1.546eV. Again,
the above parameters correspond to the low-temperature regime. The
laser light polarization is taken as
$\tilde{\boldsymbol{E}}_0\parallel\boldsymbol{x}(\boldsymbol{y})$.
In this case, we see  lineshapes with quite different Fano
parameters, compared to those for the colloidal SQD-MNP molecule.
The minimum becomes very deep for two reasons: 1) Large background
dielectric constant and, therefore, strong enhancement of electric
fields inside the system and 2) the intrinsic exciton linewidth
$1/T_{20}$ is relatively small and, therefore, the exciton-plasmon
interaction becomes ``concentrated" in a narrow energy interval. The
above reasons (1 and 2) lead to a strong interference effect for the
case shown in Fig.~\ref{epitaxial}.

It is interesting that the visibility of the exciton in SQD-MNP
molecules becomes greatly increased for small $R_d$ (Figs.
\ref{colloidal} and \ref{epitaxial}). This appears due to both the
interference effect and plasmon enhancement. The plasmon
enhancement effect is basically an increase of actual electric
fields inside the system in the regime of exciton-plasmon and
photon-plasmon resonances ($\omega_0\approx\omega_{plasmon}$ and
$\omega\approx\omega_{plasmon}$) \cite{Lakowicz,Lee,Govorov}.
Mathematically, the plasmon enhancement appears in our equations
through the factor $\frac{\epsilon_e}{\epsilon_m(\omega)}$. The
typical enhancement factors for Au-based complexes with small
$R_d$ are about 10 \cite{Govorov,Govorov2}. The increased
visibility of the exciton resonance in a SQD-MNP molecule was also
found for the dipole regime of interaction in Ref.
\cite{Zhang2006}. However our present theory provides us with an
exact solution for the electric fields and allows us to describe
this effect for the most important regime of the multipole
interaction. For example, the strong anti-resonance in
Fig.~\ref{colloidal}b is described by the factors:
$A/Q^0_{tot}\approx$ 0.005 and 0.02 for $R_d=$ 100nm and 12 nm,
respectively.

\begin{figure}[htp]
\includegraphics[width=8.5cm]{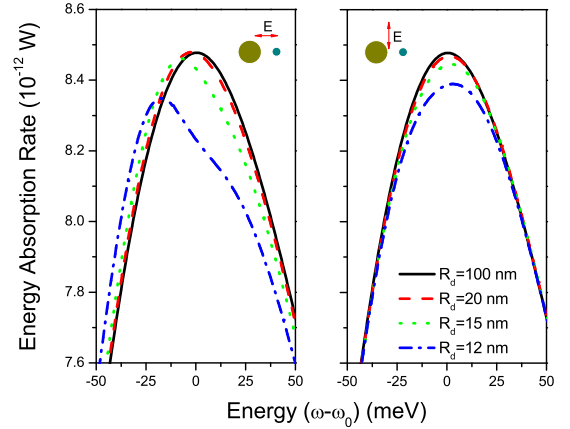}
\caption{The absorption spectrum on Dye-MNP. The dipole is parallel
(left)/ perpendicular (right) to the MNP surface}\label{dye}
\end{figure}

In the end of this section we consider the spectra of hybrid
complexes composed of optically-active molecules and MNPs. An
advantage of this system is that several or many dye molecules can
be attached to a single MNP and, therefore, the interference effect
can become much stronger \cite{Liu,Kelley}. Fig.~\ref{dye} shows the
absorption spectrum of dye-MNP(Au) for different interparticle
distances. The parameters correspond to the room temperature regime
and water environment: $1/T_{20}$=23 meV, $T_1=5$ ns,
$\varepsilon_e=1.8\varepsilon_0$ (water). Other parameters are
$R_0=10$ nm, $\mu=er_0$, $r_0=0.6$ nm, $\hbar \omega_0$=2.345eV.
Here we see that the environment has an important impact on the
energy absorption. The very large off-diagonal decoherence leads
again to quite different absorption line shapes. Importantly, even
for the large room-temperature broadening (23meV), we see the
ani-resonance. If a few or many dye molecules are attached, the
depth of the anti-resonance will increase. Neglecting the
interaction between dye molecules and the dipole orientation, we can
write $A_N=N A$, where $A_N$ is the anti-resonance coefficient for
the $N$-molecule complex, $A$ is the coefficient for a single dye
molecule, and $N$ is the number of attached dye molecules.
Correspondingly, the depth of the anti-resonance of the N-molecule
complex becomes $|\Delta Q_N|=N |\Delta Q|$, where $\Delta Q$ is
depth of the anti-resonance for a single dye molecule. From
Fig.~\ref{dye} we see that the effect of the single exciton
anti-resonance on the total absorption spectrum $|\Delta
Q|/Q_{\texttt{MNP}}^0\approx0.035$ or 3.5\%. If we now assume a
structure with $N=10$, we obtain $|\Delta
Q_N|/Q_{\texttt{MNP}}^0\approx$ 35\%. This tells us that the change
in the total absorption spectrum due to the exciton-plasmon
anti-resonance can be seen experimentally.

\section{conclusion}
We investigate the optical properties of hybrid molecules composed
of SQD, dye, and MNPs. The main focus of the paper is the regime
of strong exciton-plasmon interaction and multipole effects. First
we derive an exact analytical solution for electric fields and
absorption spectra. Then, we show that the multipole effects play
the crucial role for the strong interaction regime. When the
interparticle distance is relatively small, the multipole effect
gives significantly larger peak shifts and broadenings for the
energy absorption rate and the dipole approximation fails. The
results obtained in this paper can be used to analyze optical
experiments on hybrid systems with a strong exciton-plasmon
interaction.

\vskip 0.5 cm
\section{Acknowledgement}
This work is supported in part by the National Natural Science of
China under No. 10574017, 10774016 and a grant of the China Academy
of Engineering and Physics.  Alexander O. Govorov thanks NSF and
BNNT Initiative at Ohio University for financial support.

{\bf Electronic addresses for correspondence:}

*Electronic address: zhang$\_$wei@iapcm.ac.cn

**Electronic address: govorov@helios.phy.ohiou.edu.


\begin{thebibliography}{99}

\bibitem{Cui2001} Y. Cui, Q. Wei, H. Park, and C. M. Lieber,
Science, {\bf 293}, 1289 (2001).

\bibitem{Yin2005} Y. Yin, C. Erdonmez, and A. P. Alivisatos,
Nature, {\bf{437}}, 664 (2005).

\bibitem{Lakowicz} C. D. Geddes and J. R. Lakowicz, J. of Fluorescence, {\bf 12},
121 (2002).

\bibitem{Shimizu} K. T. Shimizu, W. K. Woo, B. R.
Fisher, H. J. Eisler, and M. G. Bawendi, Phys. Rev. Lett. 89,
117401 (2002).

\bibitem{Anger} P. Anger, P. Bharadwaj, and L. Novotny, Phys. Rev. Lett. 96, 113002 (2006).

\bibitem{Lee} J. Lee, A. O. Govorov, J. Dulka, and N. A. Kotov,
Nano Letters, 4, 2323 (2004); J. Lee, T. Javed, T. Skeini,
A. O. Govorov, G.W. Bryant, N. A. Kotov,  Angew. Chem. 45, 4819 (2006).

\bibitem{Liu} N. G. Liu, B. S. Prall, and V. I. Klimov, JACS, 128, 15362 (2006).

\bibitem{Edwards} E.W. Edwards, D. Y. Wang, H. Mohwald, Macromolecular Chemistry
and Physics, 208 439 (2007) ; G. Lu, H. Shen, B. Cheng, Z. Chen, C. A. Marquette,
L. J. Blum, O. Tillement, S. Roux, G. Ledoux, M. Ou, and P. Perriat, Appl. Phys. Lett. 89, 223128 (2006).

\bibitem{Ito} Y. Ito, K. Matsuda, and Y. Kanemitsu Y, Phys. Rev. B 75 033309 (2007) ;
V. K. Komarala, Y. P. Rakovich, A. L. Bradley, S. J. Byrne, Y. K.
Gunko, N. Gaponik, and A. Eychmller, Appl. Phys. Lett. 89, 253118
(2006).

\bibitem{Slocik} J. M. Slocik, A. O. Govorov, and R. R. Naik, Supramolecular Chem., 18 ,  415 (2006).

\bibitem{Govorov} A. O. Govorov, G. W. Bryant, W. Zhang, T. Skeini, J. Lee, N. A. Kotov,
J. M. Slocik, and R. R. Naik, Nano Lett. 6, 984 (2006).

\bibitem{Persson} B. N. J. Persson and N. D. Lang, Phys. Rev. B 26, 5409 (1982).

\bibitem{Dung} H. T. Dung, L. Knll, and D.-G. Welsch, Phys. Rev. A 62, 053804
 (2000); G.S. Agarwal and S.V. ONeil, Phys. Rev. B 28, 487 (1983)

\bibitem{Klimov} V.V. Klimov, M. Ducloy, and V.S. Letokhov, Phys. Rev. A 59, 2996 (1999).

\bibitem{Foster} Th. F$\mathbf{\ddot{o}}$rster, in Modern Quantum Chemistry, edited by O. Sinanoglu (Academic, New York,
1965).

\bibitem{Zhang2006} W. Zhang, A. O. Govorov, and G. W. Bryant,
Phys. Rev. Lett., {\bf{97}}, 146804 (2006).

\bibitem{Kelley} A. M. Kelley, Nano Lett. 7, 3235 (2007).

\bibitem{Neuhauser} D. Neuhauser and K. Lopata, J. Chem. Phys. 127, 154715 (2007).

\bibitem{Cheng}. M.-T. Cheng, S.-D. Liu, H.-J. Zhou, Z.-H. Hao, and
Q.-Q. Wang, Optics Lett., 32, 2125 (2007).

\bibitem{Fano} U. Fano, Phys. Rev. 124, 1866 (1961).

\bibitem{Driscoll} D. C. Driscoll, M. P. Hanson, A. C. Gossard, and E. R. Brown,
Appl. Phys. Lett. 86, 051908 (2005).


\bibitem{Woggon} U. Woggon et al., Nano Lett. 5, 483 (2005).

\bibitem{Mazur} Yu. I. Mazur, W. Q. Ma, X. Wang, Z. M. Wang, G. J.
Salamo, M. Xiao, T. D. Mishima, and M. B. Johnson, Appl. Phys. Lett.
83, 987 (2003)

\bibitem{Mie1908} G. Mie, Ann. Phys. {\bf{25}}, 377 (1908).

\bibitem{Kelly2003} K. L. Kelly, E. Coronado, L. L. Zhao, and G. C. Schatz,
J. Phys. Chem. B, {\bf{107}}, 668 (2003).

\bibitem{Johnson1972} P. B. Johnson, and R. W. Christy,
Phys. Rev. B, {\bf{6}}, 4370 (1972).

\bibitem{Govorov2} A. O. Govorov, and I. Carmeli,
Nano Lett., {\bf{7}}, 620 (2007).


\end{thebibliography}
\end{document}